\documentclass[sigplan,screen]{acmart}

\AtBeginDocument{%
  \providecommand\BibTeX{{%
    \normalfont B\kern-0.5em{\scshape i\kern-0.25em b}\kern-0.8em\TeX}}}

\acmConference[ICAIF-23]{November 28--29,
  2023}{New York City, NY}

\usepackage{graphicx}
\usepackage{tabularx}

\begin{document}

\title{Large Language Models in Finance: A Survey}

\author{Yinheng Li}
\authornote{All authors contributed equally to this research. Order is random}
\email{yl4039@columbia.edu}
\affiliation{%
  \institution{Columbia University}
  \streetaddress{116th and Broadway}
  \city{New York}
  \state{NY}
  \country{USA}
  \postcode{10027}
}

\author{Shaofei Wang}
\authornotemark[1]
\email{sw3316@columbia.edu}
\affiliation{%
  \institution{Columbia University}
  \streetaddress{116th and Broadway}
  \city{New York}
  \state{NY}
  \country{USA}
  \postcode{10027}
}

\author{Han Ding}
\authornotemark[1]
\email{hd2412@columbia.edu}
\affiliation{%
  \institution{Columbia University}
  \streetaddress{116th and Broadway}
  \city{New York}
  \state{NY}
  \country{USA}
  \postcode{10027}
}

\author{Hang Chen}
\authornotemark[1]
\email{hc2798@nyu.edu}
\affiliation{%
  \institution{New York University}
  \streetaddress{6 MetroTech Center}
  \city{New York}
  \state{NY}
  \country{USA}
  \postcode{11201}
}

\begin{abstract}

Recent advances in large language models (LLMs) have opened new possibilities for artificial intelligence applications in finance. In this paper, we provide a practical survey focused on two key aspects of utilizing LLMs for financial tasks: existing solutions and guidance for adoption.

First, we review current approaches employing LLMs in finance, including leveraging pretrained models via zero-shot or few-shot learning, fine-tuning on domain-specific data, and training custom LLMs from scratch. We summarize key models and evaluate their performance improvements on financial natural language processing tasks.

Second, we propose a decision framework to guide financial professionals in selecting the appropriate LLM solution based on their use case constraints around data, compute, and performance needs. The framework provides a pathway from lightweight experimentation to heavy investment in customized LLMs.

Lastly, we discuss limitations and challenges around leveraging LLMs in financial applications. Overall, this survey aims to synthesize the state-of-the-art and provide a roadmap for responsibly applying LLMs to advance financial AI.

\end{abstract}




\settopmatter{printacmref=false}

\renewcommand\footnotetextcopyrightpermission[1]{}

\maketitle

\section{Introduction}
Recent advances in artificial intelligence, especially in natural language processing, have led to the development of powerful large language models (LLMs) like ChatGPT\cite{openai2023gpt4}. These models have demonstrated impressive capabilities in understanding, generating, and reasoning about natural language. The finance industry could benefit from applying LLMs, as effective language understanding and generation can inform trading, risk modeling, customer service, and more.

In this survey, we aim to provide a practical overview focused on two key aspects of utilizing LLMs for financial applications:
\begin{itemize}
\item Existing solutions and models that employ LLMs for various finance tasks. We summarize key techniques like finetuning pretrained LLMs and training domain-specific LLMs from scratch.
\item Guidance on the decision process for applying LLMs in finance. We discuss factors to consider regarding whether LLMs are suitable for a task, cost/benefit tradeoffs, risks, and limitations.
\end{itemize}

By reviewing current literature and developments, we hope to give an accessible synthesis of the state-of-the-art along with considerations for adopting LLMs in finance. This survey targets financial professionals and researchers exploring the intersection of AI and finance. It may also inform developers applying LLM solutions for the finance industry. The remainder of the paper is organized as follows. Section 2 covers background on language modeling and recent advances leading to LLMs. Section 3 surveys current AI applications in finance and the potential for LLMs to advance in these areas. Sections 4 and 5 provide LLM solutions and decision guidance for financial applications. Finally, Sections 6 and 7 discuss risks, limitations, and conclusions.

\section{Basics of Language Models}
A language model is a statistical model that is trained on extensive text corpora to predict the probability distribution of word sequences \cite{bengio2000neural}. Let's consider a sequence of words denoted as $W = {w_1, w_2, ..., w_n}$, where $w_i$ represents the $i$-th word in the sequence. The goal of a language model is to calculate the probability $P(W)$, which can be expressed as:
\begin{eqnarray*}
P(W) &=& P(w_1, w_2, ..., w_n) \\
&=& P(w_1)P(w_2 | w_1)P(w_3 | w_1, w_2) \\
&&... P(w_n | w_1, w_2, ..., w_{n-1})
\end{eqnarray*}
The conditional probability $P(w_i | w_1, w_2, ..., w_{i-1})$ captures the likelihood of word $w_i$ given the preceding words. Over the past few decades, language model architectures have undergone significant evolution. Initially, n-gram models represented word sequences as Markov processes \cite{almutiri2022markov}, assuming that the probability of the next word depends solely on the preceding $(n-1)$ words. For example, in a bigram model, the probability of a word is only conditioned on the previous word.

Later, Recurrent Neural Network (RNN)-based models like LSTM \cite{graves2014generating} and GRU \cite{cho2014learning} emerged as neural network solutions, which are capable of capturing long-term dependencies in sequential data. However, in 2017, the introduction of the transformer architecture \cite{vaswani2017attention} revolutionized language modeling, surpassing the performance of RNNs in tasks such as machine translation. Transformers employ self-attention mechanisms to model parallel relationships between words, facilitating efficient training on large-scale datasets. Prominent transformer-based models include GPT (Generative Pre-trained Transformer) \cite{brown2020language, wang2020generalizing}, which is decoder-only framework, BERT (Bidirectional Encoder Representations from Transformers) \cite{devlin2019bert}, which is encoder-only framework, and T5 (Text-to-Text Transfer Transformer) \cite{raffel2020exploring}, which leverages both encoder and decoder structures. These models have achieved state-of-the-art results on various natural language processing (NLP) tasks through transfer learning.

It is important to note that the evolution of language models has mainly been driven by advancements in computational power, the availability of large-scale datasets, and the development of novel neural network architectures. These models have significantly enhanced language understanding and generation capabilities, enabling their application across a wide range of industries and domains.

\section{Overview of AI Applications in Finance}

\subsection{Current AI Applications in Finance}

Artificial Intelligence (AI) has witnessed extensive adoption across various domains of finance in recent years \cite{AIinfinance}. In this survey, we focus on key financial applications, including trading and portfolio management \cite{Zhang_2020}, financial risk modeling \cite{9249416}, financial text mining \cite{fintextmining, 10.1145/3490354.3494388}, and financial advisory and customer services \cite{Shah_Raj_Kumar_P_H}. While this list is not exhaustive, these areas have shown significant interest and high potential with the advancement of AI.

\textbf{Trading and portfolio management} have been early adopters of machine learning and deep learning models within the finance industry. The primary objective of trading is to forecast prices and generate profits based on these predictions. Initially, statistical machine learning methods such as Support Vector Machines (SVM) \cite{KIM2003307}, Xgboost \cite{zolotareva2021aiding}, and tree-based algorithms were utilized for profit and loss estimation. However, the emergence of deep neural networks introduced techniques like Recurrent Neural Networks (RNN), particularly Long Short-Term Memory (LSTM) networks \cite{deepnntrading}, Convolutional Neural Networks (CNN), and transformers \cite{wen2023transformers}, which have proven effective in price forecasting. Additionally, reinforcement learning \cite{wang2019dynamic} has been applied to automatic trading and portfolio optimization.

\textbf{Financial risk modeling} encompasses various applications of machine learning and deep learning models. For instance, McKinsey \& Company has developed a deep learning-based solution for financial fraud detection by leveraging user history data and real-time transaction data \cite{deeplearning_fraud_detection}. Similar approaches have been employed in credit scoring \cite{creditscoring_deeplearning, creditscorenn} and bankruptcy or default prediction \cite{bankruptcyprediction}.

\textbf{Financial text mining} represents a popular area where deep learning models and natural language processing techniques are extensively utilized. According to \cite{deeplearninginfinancesurvcey}, there are over 40 research publications on this topic. Financial text mining aims to extract valuable information from large-scale unstructured data in real-time, enabling more informed decision-making in trading and risk modeling. For example, \cite{trading_Sentiment} employs financial market sentiment extracted from news articles to forecast the direction of the stock market index.

\textbf{Applying AI in financial advisory and customer-related services} is an emerging and rapidly growing field. AI-powered chatbots, as discussed in \cite{chatbot}, already provide more than 37\% of supporting functions in various e-commerce and e-service scenarios. In the financial industry, chatbots are being adopted as cost-effective alternatives to human customer service, as highlighted in the report "Chatbots in consumer finance" \cite{Chatbotsinconsumerfinance}. Additionally, banks like JPMorgan are leveraging AI services to provide investment advice, as mentioned in a report by CNBC \cite{jpmorgan}.

The current implementation of deep learning models offers significant advantages by efficiently extracting valuable insights from vast amounts of data within short time frames. This capability is particularly valuable in the finance industry, where timely and accurate information plays a crucial role in decision-making processes. With the emergence of LLMs, even more tasks that were previously considered intractable become possible, further expanding the potential applications of AI in the finance industry.

\subsection{Advancements of LLMs in Finance}

LLMs offer numerous advantages over traditional models, particularly in the field of finance. Firstly, LLMs leverage their extensive pre-training data to effectively process common-sense knowledge, enabling them to understand natural language instructions. This is valuable in scenarios where supervised training is challenging due to limited labeled financial data or restricted access to certain documents. LLMs can perform tasks through zero-shot learning \cite{zeroshot}, as demonstrated by their satisfactory performance in sentiment classification tasks across complex levels \cite{zhang2023sentiment}. For similar text mining tasks on financial documents, LLMs can automatically achieve acceptable performance.

Compared to other supervised models, LLMs offer superior adaptation and flexibility. Instead of training separate models for specific tasks, LLMs can handle multiple tasks by simply modifying the prompt under different task instructions \cite{gpt3}. This adaptability does not require additional training, enabling LLMs to simultaneously perform sentiment analysis, summarization, and keyword extraction on financial documents.

LLMs excel at breaking down ambiguous or complex tasks into actionable plans. Applications like Auto-GPT \cite{autogpt}, Semantic Kernel \cite{semantic_kernel}, and LangChain \cite{Chase_LangChain_2022} have been developed to showcase this capability. In this paper, we refer to this as \textbf{Tool Augmented Generation}. For instance \cite{AutoGPTforfinance}, Auto-GPT can optimize a portfolio with global equity ETFs and bond ETFs based on user-defined goals. It formulates detailed plans, including acquiring financial data, utilizing Python packages for Sharpe ratio optimization, and presenting the results to the user. Previously, achieving such end-to-end solutions with a single model was unfeasible. This property makes LLMs an ideal fit for financial customer service or financial advisory, where they can understand natural language instructions and assist customers by leveraging available tools and information.

While the application of LLMs in finance is really promising, it is crucial to acknowledge their limitations and associated risks, which will be further discussed in Section 6.

\section{LLM Solutions for Finance}

\subsection{Utilizing Few-shot/Zero-shot Learning in Finance Applications}

Accessing LLM solutions in finance can be done through two options: utilizing an API from LLM service providers or employing open-source LLMs. Companies like OpenAI\footnote{https://openai.com/product}, Google\footnote{https://bard.google.com/}, and Microsoft\footnote{https://azure.microsoft.com/en-us/products/cognitive-services/openai-service} offer LLM services through APIs. These services not only provide the base language model capabilities but also offer additional features tailored for specific use cases. For example, OpenAI's APIs include functionalities for chat, SQL generation, code completion, and code interpretation. While there is no dedicated LLM service exclusively designed for finance applications, leveraging these general-purpose LLM services can be a viable option, especially for common tasks. An example in this work \cite{mlq-ai} demonstrates the use of OpenAI's GPT4 service for financial statement analysis.

In addition to LLM services provided by tech companies, open-source LLMs can also be applied to financial applications. Models such as LLaMA \cite{touvron2023llama}, BLOOM \cite{workshop2023bloom}, Flan-T5 \cite{chung2022scaling}, and more are available for download from the Hugging Face model repository\footnote{https://huggingface.co/models}. Unlike using APIs, hosting and running these open-source models would require self-hosting. Similar to using LLM APIs, zero-shot or few-shot learning approaches can be employed with open-source models. Utilizing open-source models offers greater flexibility as the model's weights are accessible, and the model's output can be customized for downstream tasks. Additionally, it provides better privacy protection as the model and data remain under user's control. However, working with open-source models also has its drawbacks. Reported evaluation metrics suggest a performance gap between open-source models and proprietary models. For certain downstream tasks, zero-shot or few-shot learning may not yield optimal performance. In such cases, fine-tuning the model with labeled data, expertise, and computational resources is necessary to achieve satisfactory results. This may explain why, at the time of writing this paper, no direct examples of open-source models applied to financial applications have been found. In Section 5, we provide a more detailed discussion of which option is more favorable under different circumstances.

\begin{table*}[htbp]
  \centering
  \caption{Quick Overview of Finetuned Finance LLM}
  \label{tab:plm-sizes}
  \begin{tabular}{lllll}
    \toprule
    \textbf{Model Name} & \textbf{Finetune data size (samples)} & \textbf{Training budget} & \textbf{Model architecture} & \textbf{Release time} \\
    \midrule
    FinMA-7B & Raw: 70k, Instruction: 136k & 8 A100 40GB GPUs & LLaMA-7B & Jun 2023 \\
    FinMA-30B & Raw: 70k, Instruction: 136k & 128 A100 40GB GPUs & LLaMA-30B & Jun 2023 \\
    Fin-GPT(V1/V2/V3) & 50K & $< \$300$ per training & ChatGLM, LLaMA & July 2023 \\
    Instruct-FinGPT & 10K Instruction & 8 A100 40GB GPUs, $\sim$1 hr & LLaMA-7B & Jun 2023 \\
    Fin-LLaMA\cite{Fin-LLAMA} & 16.9K Instruction & NA & LLaMA-33B & Jun 2023 \\
    Cornucopia(Chinese)\cite{Cornucopia-LLaMA-Fin-Chinese} & 12M instruction & NA & LLaMA-7B & Jun 2023 \\
    \bottomrule
  \end{tabular}
\end{table*}
 
\begin{table*}[ht]
  \centering
  \caption{Quick Overview of from scratch trained Finance LLMs}
  \label{tab:finance-llms}
  \begin{tabular}{p{2cm} p{4cm} p{3cm} p{4cm} p{2cm}}
    \toprule
    \textbf{Pretrained LLM} & \textbf{Corpus size(tokens)} & \textbf{Training budget(A100·hours)} & \textbf{Model architecture}  & \textbf{Release time} \\
    \midrule
    BloomBergGPT & 363B Finance tokens + 345B public tokens & 1,300,000 &  50B-BLOOM & May 2023 \\
    XuanYuan2.0 & 366B for pre-training + 13B for finetuning & Not released & 176B-BLOOM & May 2023 \\
    Fin-T5 & 80B Finance tokens & Days/weeks & 770M-T5  & Feb 2023\\
    \bottomrule
  \end{tabular}
\end{table*}

\subsection{Fine-tuning a Model}

Fine-tuning LLMs in the finance domain can enhance domain-specific language understanding and contextual comprehension, resulting in improved performance in finance-related tasks and generating more accurate and tailored outputs.

\subsubsection{Common Techniques for LLM Fine-tuning}

Modern techniques for fine-tuning LLMs typically fall into two main categories: standard fine-tuning and instructional fine-tuning.

In standard fine-tuning, the model is trained on the raw datasets without modification. The key context, question, and desired answer are directly fed into the LLM, with the answer masked during training so that the model learns to generate it. Despite its simplicity, this approach is widely effective.

Instruct fine-tuning \cite{instruct} involves creating task-specific datasets that provide examples and guidance to steer the model's learning process. By formulating explicit instructions and demonstrations in the training data, the model can be optimized to excel at certain tasks or produce more contextually relevant and desired outputs. The instructions act as a form of supervision to shape the model's behavior.

Both methods have their merits: standard fine-tuning is straightforward to implement, while instructional fine-tuning allows for more precise guidance of the model. The ideal approach depends on the amount of training data available and the complexity of the desired behaviors. However, both leverage the knowledge already embedded in LLMs and fine-tune them for enhanced performance on downstream tasks.

In addition to the above methods, techniques such as Low-Rank Adaptation (LoRA)\cite{hu2021lora} and quantization\cite{gholami2021survey} can enable fine-tuning with significantly lower computational requirements.

LoRA allows for fine-tuning the low-rank decomposed factors of the original weight matrices instead of the full matrices. This approach drastically reduces the number of trainable parameters, enabling training on less powerful hardware and shortening the total training time.

Another impactful approach is to use reduced numerical precisions such as bfloat16 \cite{kalamkar2019study} or float16 instead of float32. By halving the bit-width, each parameter only occupies 2 bytes instead of 4 bytes, reducing memory usage by 50\%. This also accelerates computation by up to 2x since smaller data types speed up training. Moreover, the reduced memory footprint enables larger batch sizes, further boosting throughput.

\subsubsection{Fine-tuned finance LLM evaluation}

The performance of fine-tuned finance LLMs can be evaluated in two categories: finance classification tasks and finance generative tasks. In finance classification, we consider tasks such as Sentiment Analysis and News Headline Classification. In finance generative tasks, our focus is on Question Answering, News Summarization, and Named Entity Recognition. Table 1 provides detailed information about all the fine-tuned finance LLMs. Among the various fine-tuned LLMs, we will focus on discussing three of them: (1) PIXIU (also known as FinMA)\cite{xie2023pixiu}, fine-tuned LLaMA on 136K task-specific instruction samples. (2) FinGPT\cite{yang2023fingpt}, it presents a end-to-end framework for training and applying FinLLMs in the finance industry. FinGPT utilizes the lightweight Low-rank Adaptation (LoRA) technique to fine-tune open-source LLMs (such as LLaMA and ChatGLM) using approximately 50k samples. However, FinGPT's evaluation is only limited to finance classification tasks. (3) Instruct-FinGPT\cite{zhang2023instructfingpt}, on the other hand, fine-tunes LLaMA on 10k instruction samples derived from two Financial Sentiment Analysis Datasets and also solely evaluates performance on finance classification tasks.

Based on the reported model performance, we summarize our findings as below:
\begin{itemize}
\item Compared to the original base LLM (LLaMA) and other open-source LLMs (BLOOM, OPT\cite{zhang2022opt}, ChatGLM\cite{zeng2023glm-130b, du2022glm}), all fine-tuned finance LLMs exhibit significantly better performance across all finance-domain tasks reported in the papers, especially classification tasks.
\item The fine-tuned finance LLMs outperform BloombergGPT\cite{wu2023bloomberggpt} in most finance tasks reported in the papers. 
\item When compared to powerful general LLMs like ChatGPT and GPT-4, the fine-tuned finance LLMs demonstrate superior performance in most finance classification tasks, which indicates their enhanced domain-specific language understanding and contextual comprehension abilities. However, in finance generative tasks, the fine-tuned LLMs show similar or worse performance, suggesting the need for more high-quality domain-specific datasets to improve their generative capabilities.
\end{itemize}

\subsection{Pretrain from Scratch}
The objective of training LLMs from scratch is to develop models that have even better adaptation to the finance domain. Table 2 presents the current finance LLMs that have been trained from scratch: BloombergGPT, Xuan Yuan 2.0 \cite{zhang2023xuanyuan}, and Fin-T5\cite{lu2023bbtfin}.

As shown in Table 2, there is a trend of combining public datasets with finance-specific datasets during the pretraining phase. Notably, BloombergGPT serves as an example where the corpus comprises an equal mix of general and finance-related text. It is worth mentioning that BloombergGPT primarily relies on a subset of 5 billion tokens that pertain exclusively to Bloomberg, representing only 0.7\% of the total training corpus. This targeted corpus contributes to the performance improvements achieved in finance benchmarks.

Both BloombergGPT and Fin-T5 have demonstrated superior performance compared to their original models like BLOOM176B and T5, respectively. These tasks encompass activities such as market sentiment classification, multi-categorical and multi-label classification, and more. BloombergGPT achieves an impressive average score of 62.51, surpassing the open-source BLOOM176B model, which only attains a score of 54.35. Similarly, Fin-T5 demonstrates its excellence with an average score of 81.78, outperforming the T5 model's score of 79.56. Notably, BloombergGPT was evaluated using an internal benchmark specifically designed by Bloomberg. The results of this evaluation showcased remarkable improvements, as BloombergGPT achieved an average score of 62.47, surpassing the performance of BLOOM176B, which only attained a score of 33.39. This outcome highlights that even when the internal private training corpus constitutes less than 1\% of the total training corpus, it can still lead to substantial enhancements in evaluating tasks within the same domain and distribution.

On finance-related generative tasks such as Question Answering, Named Entity Recognition, summarization, both models exhibited significantly better results compared to their respective general models by a considerable margin. Specifically, BloombergGPT achieved an impressive score of 64.83, surpassing BLOOM-176B's score of 45.43. Similarly, Fin-T5 outperformed T5 with a score of 68.69, while T5 scored 66.06. These findings further highlight the models' superior performance in generating finance-related content when compared to their general-purpose counterparts.

Although these models are not as powerful as closed-source models like GPT-3 or PaLM\cite{chowdhery2022palm}, they demonstrate similar or superior performance compared to similar-sized public models. In evaluations on various general generative tasks, such as BIG-bench Hard, knowledge assessments, reading comprehension, and linguistic tasks, BloombergGPT exhibited comparable or superior performance compared to similar-sized public models, albeit slightly inferior to larger models like GPT-3 or PaLM. Overall, BloombergGPT showcased commendable performance across a wide range of general generative tasks, positioning it favorably among models of comparable size. This indicates that the model's enhanced capabilities in finance-related tasks do not come at the expense of its general abilities.

\section{Decision Process in Applying LLM to Financial Applications}

\subsection{Determining the Need for a LLM}

Before exploring LLM solutions, it is essential to ascertain whether employing such a model is truly necessary for the given task. The advantages of LLMs over smaller models can be summarized as follows, as outlined in the work by Yang et al. \cite{yang2023harnessing}:

\textbf{Leveraging Pretraining Knowledge:} LLMs can utilize the knowledge acquired from pretraining data to provide solutions. If a task lacks sufficient training data or annotated data but requires common-sense knowledge, an LLM may be a suitable choice.

\textbf{Reasoning and Emergent Abilities:} LLMs excel at tasks that involve reasoning or emergent abilities \cite{wei2022emergent}. This property makes LLMs well-suited for tasks where task instructions or expected answers are not clearly defined, or when dealing with out-of-distribution data. In the context of financial advisory, client requests in customer service often exhibit high variance and complex conversations. LLMs can serve as virtual agents to provide assistance in such cases.

\textbf{Orchestrating Model Collaboration:} LLMs can act as orchestrators between different models and tools. For tasks that require collaboration among various models, LLMs can serve as orchestrators to integrate and utilize these tools together \cite{Chase_LangChain_2022, autogpt, semantic_kernel}. This capability is particularly valuable when aiming for a robust automation of a model solution pipeline.

While LLMs offer immense power, their use comes with a significant cost, whether utilizing a third-party API \cite{openai2023gpt4} or fine-tuning an open-source LLM. Therefore, it is prudent to consider conventional models before fully committing to LLMs. In cases where the task has a clear definition (e.g., regression, classification, ranking), there is an ample amount of annotated training data, or the task relies minimally on common-sense knowledge or emerging capabilities like reasoning, relying on LLMs may not be necessary or justified initially.

\subsection{A general decision guidance for applying LLMs on finance tasks}
\begin{figure*}[htp]
    \centering
    \includegraphics[width=\textwidth]{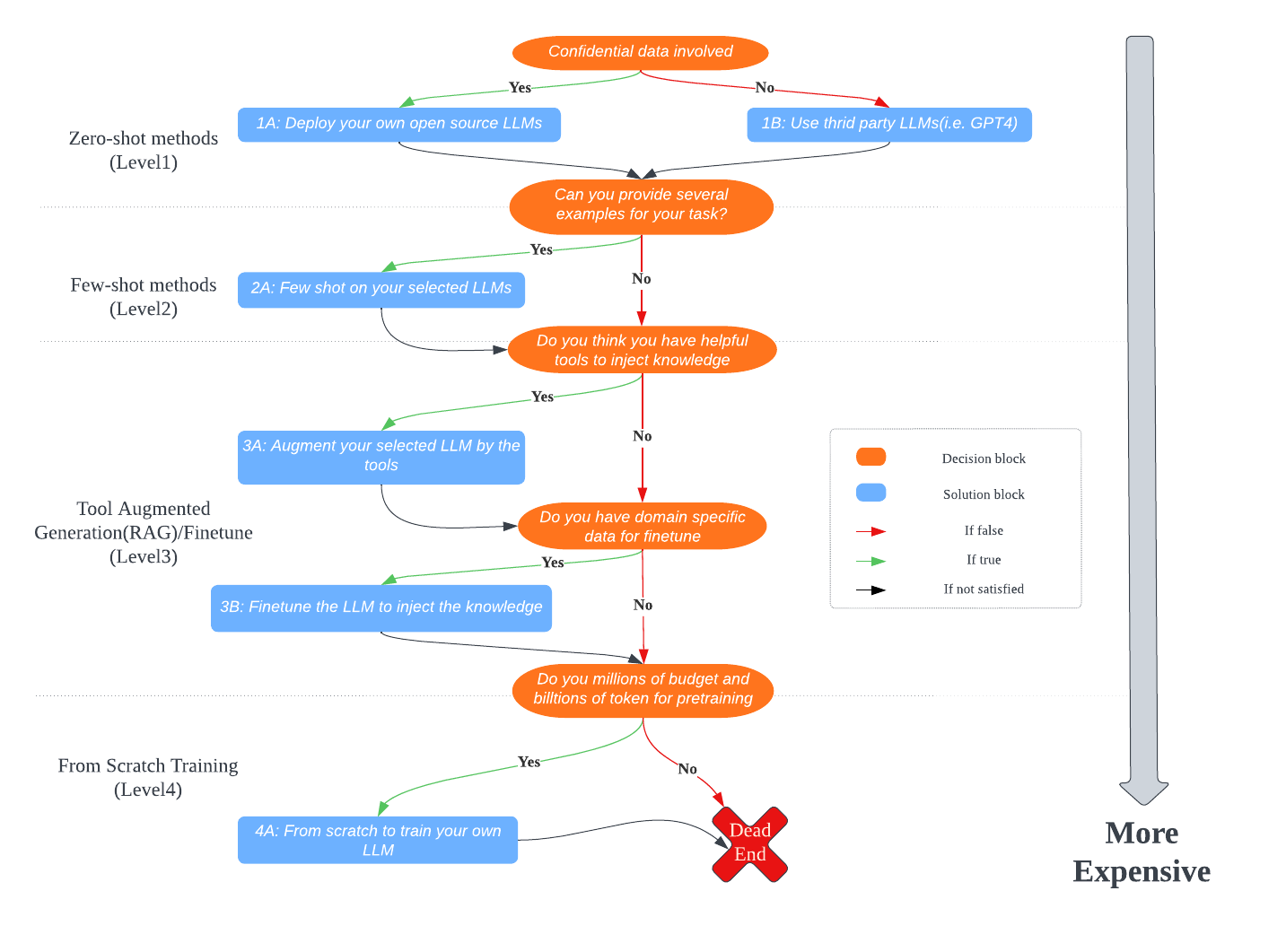}
    \caption{Decision process flow chart}
    \label{fig:galaxy}
\end{figure*}

\begin{table*}[htbp]
  \centering
  \label{tab:cost}
  \begin{tabular}{p{4.5cm} p{3cm} p{3cm} p{3.5cm}}
    \toprule
    \textbf{Options} & \textbf{Development Computational Cost(\$)} & \textbf{Development Data Cost(samples)} & \textbf{Deployment Computational Cost (\$/1k tokens generated)} \\
    \midrule
    OpenSource-ZeroShot & - & - & 0.006 - 0.037 \\
    3rd party-ZeroShot & - & - & 0.002 - 0.12 \\
    \midrule
    OpenSource-FewShot & - & - & 0.006 - 0.037 \\
    3rd party-FewShot & - & - & 0.002 - 0.12 \\
    \midrule
    OpenSource Tool Augmented Generation & Cost of developing tools & - & 0.006 - 0.037 \\
    3rd party Tool Augmented Generation & Cost of developing tools & - & 0.002 - 0.12 \\
    \midrule
    OpenSource-Finetune & 4-360,000 & 10,000 - 12,000,000 & 0.0016 - 0.12 \\
    3rd party-Finetune & 30-30,000 & 10,000 - 12,000,000 & 0.002 - 0.12 \\
    \midrule
    Train from Scratch & 5,000,000 & 700,000,000 & 0.0016 - 0.12 \\
    \bottomrule
  \end{tabular}
  \caption{Costs of Different LLM Options: This table gives an approximate range of requirements of data and dollar cost. The data and dollar cost requirements for development are estimated based on previous works listed in \ref{tab:finance-llms}. The third-party deployment costs are listed in https://openai.com/pricing. The open source deployment costs are calculated based on https://openai.com/pricing and https://aws.amazon.com/ec2/pricing/on-demand/. We assume using NVIDIA A100 GPU. The cost of \$/ tokens = \$ / second * second / 1k tokens, and it typically takes 3 to 33 seconds to generate 1k tokens, depending on model size.}
\end{table*}

Once the decision has been made to utilize LLMs for a finance task, a decision guidance framework can be followed to ensure efficient and effective implementation. The framework, illustrated in Figure \ref{fig:galaxy}, categorizes the usage of LLMs into four levels based on computational resources and data requirements. By progressing through the levels, costs associated with training and data collection increase. It is recommended to start at Level 1 and move to higher levels (2, 3, and 4) only if the model's performance is not satisfactory. The following section provides detailed explanations of the decision and action blocks at each level. Table \ref{tab:cost} presents an approximate cost range for different options, based on pricing from various third-party services like AWS and OpenAI.

\subsubsection{Level 1: Zero-shot Applications}
The first decision block determines whether to use an existing LLM service or an open-source model. If the input question or context involves confidential data, it is necessary to proceed with the 1A action block, which involves self hosting an open-source LLM. As of July 2023, several options are available, including LLAMA\cite{touvron2023llama}, OpenLLAMA\cite{openlm2023openllama}, Alpaca\cite{alpaca}, and Vicuna\cite{vicuna2023}. LLAMA offers models with sizes ranging from 7B to 65B, but they are limited to research purposes. OpenLLAMA provides options for 3B, 7B, and 13B models, with support for commercial usage. Alpaca and Vicuna are fine-tuned based on LLAMA, offering 7B and 13B options. Deploying your own LLM requires a robust local machine with a suitable GPU, such as NVIDIA-V100 for a 7B model or NVIDIA-A100, A6000 for a 13B model.

If data privacy is not a concern, selecting third-party LLMs such as GPT3.5/GPT4 from OpenAI or BARD from Google is recommended. This option allows for lightweight experiments and early performance evaluation without significant deployment costs. The only cost incurred would be the fees associated with each API call, typically based on input length and the token count of the model's output.

\subsubsection{Level 2: Few-shot Applications}
If the model's performance at Level 1 is not acceptable for the application, few-shot learning can be explored if there are several example questions and their corresponding answers available. Few-shot learning has shown advantages in various previous works \cite{brown2020language, wang2020generalizing}. The core idea is to provide a set of example questions along with their corresponding answers as context in addition to the specific question being asked. The cost associated with few-shot learning is similar to that of the previous levels, except for the requirement of providing examples each time. Generally, achieving good performance may require using 1 to 10 examples. These examples can be the same across different questions or selected based on the specific question at hand. The challenge lies in determining the optimal number of examples and selecting relevant ones. This process involves experimentation and testing until the desired performance boundary is reached.

\subsubsection{Level 3: Tool-Augmented Generation and Finetuning}
If the task at hand is extremely complicated and in-context learning does not yield reasonable performance, the next option is to leverage external tools or plugins with the LLM, assuming a collection of relevant tools/plugins is available. For example, a simple calculator could assist with arithmetic-related tasks, while a search engine could be indispensable for knowledge-intensive tasks such as querying the CEO of a specific company or identifying the company with the highest market capitalization.

Integrating tools with LLMs can be achieved by providing the tool's descriptions. The cost associated with this approach is generally higher than that of few-shot learning due to the development of the tool(s) and the longer input sequence required as context. However, there may be instances where the concatenated tool description is too long, surpassing the input length limit of LLMs. In such cases, an additional step such as a simple tool retrieval or filter might be needed to narrow down the tools for selection. The deployment cost typically includes the cost of using the LLMs as well as the cost of using the tool(s).

If the above options fail to produce satisfactory performance, finetuning the LLMs can be attempted. This stage requires a reasonable amount of annotated data, computational resources (GPU, CPU, etc.), and expertise in tuning language models, as listed in Table \ref{tab:cost}.

\subsubsection{Level 4: Train Your Own LLMs from Scratch}
If the results are still unsatisfactory, the only option left is to train domain-specific LLMs from scratch, similar to what BloombergGPT did. However, this option comes with significant computational costs and data requirements. It typically requires millions of dollars in computational resources and training on a dataset with trillions of tokens. The intricacies of the training process are beyond the scope of this survey, but it is worth noting that it can take several months or even years of effort for a professional team to accomplish.

By following this decision guidance framework, financial professionals and researchers can navigate through the various levels and options, making informed choices that align with their specific needs and resource constraints.

\subsection{Evaluation}

The evaluation of LLMs in finance can be conducted through various approaches. One direct evaluation method is to assess the model's performance on downstream tasks. Evaluation metrics can be categorized into two main groups: accuracy and performance, based on the taxonomy provided by \cite{nlpfinance}. The accuracy category can further be divided into metrics for regression (such as MAPE, RMSE, $R^2$) and metrics for classification (Recall, Precision, F1 score). The performance category includes metrics or measurements that directly assess the model's performance on the specific task, such as measuring total profit or Sharpe Ratio in a trading-related task. These evaluations can be conducted using historical data, backtest simulations, or online experiments. While performance metrics are often more important in finance, it is crucial to ensure that accuracy metrics align with performance to ensure meaningful decision-making and guard against overfitting.

In addition to task-specific evaluations, general metrics used for LLMs can also be applied. Particularly, when evaluating the overall quality of an existing LLM or a fine-tuned one, comprehensive evaluation systems like the one presented in \cite{liang2022holistic} can be utilized. This evaluation system covers tasks for various scenarios and incorporates metrics from different aspects, including accuracy, fairness, robustness, bias, and more. It can serve as a guide for selecting a language model or evaluating one's own model in the context of finance applications.

\subsection{Limitations}

While significant progress has been made in applying LLMs to revolutionize financial applications, it is important to acknowledge the limitations of these language models. Two major challenges are the production of disinformation and the manifestation of biases, such as racial, gender, and religious biases, in LLMs \cite{tamkin2021understanding}. In the financial industry, accuracy of information is crucial for making sound financial decisions, and fairness is a fundamental requirement for all financial services. To ensure information accuracy and mitigate hallucination, additional measures like retrieve-augmented generation \cite{lewis2021retrievalaugmented} can be implemented. To address biases, content censoring and output restriction techniques (such as only generating answers from a pre-defined list) can be employed to control the generated content and reduce bias.

LMMs poises potential challenges in terms of regulation and governance. Although LLM offers more interpretability compared to conventional deep learning models by providing reasoning steps or thinking processes for the generated answers when prompted correctly \cite{COT} \cite{yao2023tree}, LLM remains a black box and explainability of the content it generates is highly limited.

Addressing these limitations and ensuring the ethical and responsible use of LLMs in finance applications is essential. Continuous research, development of robust evaluation frameworks, and the implementation of appropriate safeguards are vital steps in harnessing the full potential of LLMs while mitigating potential risks.

\section{Conclusion}

In conclusion, this paper has conducted a timely and practical survey on the emerging application of LLMs for financial AI. We structured the survey around two critical pillars: solutions and adoption guidance.

Under solutions, we reviewed diverse approaches to harnessing LLMs for finance, including leveraging pretrained models, fine-tuning on domain data, and training custom LLMs. Experimental results demonstrate significant performance gains over general purpose LLMs across natural language tasks like sentiment analysis, question answering, and summarization.

To provide adoption guidance, we proposed a structured framework for selecting the optimal LLM strategy based on constraints around data availability, compute resources, and performance needs. The framework aims to balance value and investment by guiding practitioners from low-cost experimentation to rigorous customization.

In summary, this survey synthesized the latest progress in applying LLMs to transform financial AI and provided a practical roadmap for adoption. We hope it serves as a useful reference for researchers and professionals exploring the intersection of LLMs and finance. As datasets and computation improve, finance-specific LLMs represent an exciting path to democratize cutting-edge NLP across the industry.

\bibliographystyle{ACM-Reference-Format}
\bibliography{sample-base}


\end{document}